\documentclass[preprint,aps]{revtex4}

\usepackage{amsmath,amssymb,amsfonts,dcolumn,color,graphicx,graphics,latexsym,placeins,epsfig}

\newcommand{\be}{\begin{eqnarray}}
\newcommand{\ee}{\end{eqnarray}}
\newcommand{\ba}{\begin{eqnarray}}
\newcommand{\ea}{\end{eqnarray}}

\begin{document}

\title{Brans-Dicke Theory with $\Lambda {\ensuremath >} 0$: Black Holes and Large Scale Structures}
\author{$^{1}$Sourav Bhattacharya\footnote{Current affiliation : Inter-University Centre for Astronomy and Astrophysics (IUCAA), Pune-411007, India}}
\email{souravbhatta@physics.uoc.gr}
\author{$^{1}$Konstantinos F. Dialektopoulos}
\email{kdialekt@physics.uoc.gr}
\author{$^{1,2}$Antonio Enea Romano}
\email{aer@physics.uoc.gr}
\author{$^{1}$Theodore N. Tomaras}
\email{tomaras@physics.uoc.gr}
\affiliation{$^{1}$ITCP and Department of Physics, 
	University of Crete, 70013 Heraklion, Greece\\}
\affiliation{$^{2}$Instituto de Fisica, Universidad de Antioquia, A. A. 1226, Medellin, Colombia}	

\date{\today}

\begin{abstract}
\noindent
A step-by-step approach is followed to study cosmic structures in the context of Brans-Dicke theory with positive cosmological constant $\Lambda$ and parameter $\omega$. First, it is shown that regular stationary black-hole solutions not only have constant Brans-Dicke field $\phi$, but can exist only for $\omega=\infty$, which forces the theory to coincide with the General Relativity. Generalizations of the theory in order to evade this black-hole no-hair theorem are presented. It is also shown that in the absence of a stationary cosmological event horizon in the asymptotic region, a stationary black hole horizon can support a non-trivial Brans-Dicke hair. 
Even more importantly, it is shown next, that the presence of a stationary cosmological event horizon rules out any regular stationary solution, appropriate for the description of a star. Thus, to describe a star one has to assume that there is no such stationary horizon in the faraway asymptotic region.  Under this implicit assumption generic spherical cosmic structures are studied perturbatively and is shown that only for $\omega>0$ or $\omega\lesssim -5$ their predicted maximum sizes are consistent with observations. We also point out how, many of the conclusions of this work differ qualitatively from the $\Lambda=0$ spacetimes.   \\
\end{abstract}


\maketitle

\maketitle
\section{Introduction}
It is well established by now, that the $\Lambda$CDM model is remarkably successful in the
interpretation of cosmological observations, including the cosmic
microwave background, supernovae type Ia, and the large scale 
structure in the Universe.  However, the up to now failure to confirm the existence of a TeV-scale supersymmetry or of any of the prime candidates for particle dark matter - despite several potentially interesting signals - has intensified the analysis of alternative proposals as
resolutions of the ``Dark Universe" puzzle.  
 
The Brans-Dicke theory~\cite{Brans} with a positive cosmological constant is the simplest among the scalar-tensor viable alternatives to Einstein gravity and is currently a subject of intensive research. In the Jordan frame the theory is~\cite{review, Avilez:2013dxa}
\begin{eqnarray}
{\cal{L}}=\sqrt{-g}\left[\phi R-2\Lambda-\frac{\omega}{\phi}\left(\nabla \phi\right)^2\right]+{\cal{L}}_{\rm M}.
\label{lg}
\end{eqnarray}
The constant $\omega$ is the Brans-Dicke parameter, while the inverse of the Brans-Dicke field $\phi$ can be interpreted as a space-time dependent Newton's ``constant". The equations of motion are 
\begin{eqnarray}
R_{ab}&=&\frac{\Lambda(2\omega+1) g_{ab}}{(2\omega+3) \phi}+\frac{T_{ab}}{\phi}-\frac{(\omega+1)T g_{ab}}{(2\omega+3)\phi}
+\frac{\omega}{\phi^2} \nabla_a\phi \nabla_b\phi + \frac{\nabla_a\nabla_b\phi}{\phi}\nonumber \\
\Box\phi&=&(T-4\Lambda)/(2\omega +3)
\label{fielded}
\end{eqnarray}
where $T$ is the trace of the energy-momentum tensor corresponding to ${\cal{L}}_{\rm M}$, 
which describes collectively all matter fields. For a review and list of references on the Brans-Dicke theory see e.g.~\cite{review}.

Our ultimate objective here is the study of cosmic structures in the context of this theory. However, as will become apparent below, the study of black-holes will give us useful insight about how to proceed. 
In the context of Einstein gravity it is known that under certain symmetry, field-boundedness and energy conditions, there can be no nontrivial real field configuration outside the event horizon of a stationary black hole, except for long range gauge fields~\cite{Bekenstein:1998aw, Bekenstein, Herdeiro}. 
 This was extended to the theory (\ref{lg}) with $\Lambda=0$, for which
it was shown in~\cite{Hawking:1972qk} that {\it asymptotically flat} stationary black holes do not support a nontrivial profile for the scalar field. This was generalized recently~\cite{Sotiriou:2011dz} to a class of scalar-tensor gravities with $\omega=\omega(\phi)$ and generic convex potential $V(\phi)$. We refer the reader to~\cite{Hui:2012qt, Graham:2014mda, Chris, Sotiriou} for such theorems and possible violations in examples of scalar tensor gravity. See also~\cite{Herdeiro} and references therein for an interesting violation of scalar no hair theorems in Einstein gravity. Precisely, by complexifying the scalar,
one can still retain the desired symmetries of the energy-momentum tensor and hence the spacetime, while having non-trivial hairy solutions.
   
All the above proofs assume $\Lambda=0$ and asymptotic flatness. One exception to this is the non-trivial black hole, known to exist in (\ref{lg}) for $\Lambda<0$ and asymptotic Lifshitz boundary condition~\cite{Maeda:2011jj}.

However, the recent discovery of the accelerated cosmic expansion, which suggests that our universe may be endowed with a positive $\Lambda \sim10^{-52}{\rm m}^{-2}$, makes it physically most interesting to study solutions of (\ref{lg}) with asymptotic de Sitter behavior. 
A crucial feature of a stationary spacetime with $\Lambda>0$ is the existence of a cosmological event horizon, a null hypersurface of length scale $\sim 1/\sqrt{\Lambda}$ acting as a causal boundary surrounding us~\cite{Gibbons:1977mu, Kastor:1992nn}. It sets a limit on the size of the observable universe, since no causal communication is possible beyond it.
Geometrically, such a horizon resembles the black hole horizon, i.e. it is a Killing horizon with vanishing expansion, shear and rotation for null geodesics tangent to it. 
Uniqueness and no-hair properties of such black holes in Einstein gravity can be found in e.g.~\cite{Boucher}-\cite{Bhattacharya:2007zzb}.

In this work we focus on theory (\ref{lg}) with $\Lambda>0$. In Sections II and III we extend to this theory and generalizations thereof the study of black holes. The obtained results are interesting in their own right, but in addition are useful for the discussion of cosmic structures, which is the content of Section IV. We show that, maybe surprisingly, no star solution can exist in (\ref{lg}) in the presence of a regular stationary cosmological event horizon. Thus, such structures can be accommodated in (\ref{lg}) only if one discards asymptotic Schwarzschild-de Sitter behavior of space-time. Finally, with this implicit assumption the maximum sizes of such structures are studied perturbatively, and from their comparison with observations constraints on the parameter $\omega$ are obtained. We end with a short discussion and outlook.  

\section{No hair in Brans-Dicke with $\Lambda>0$}
Let us begin by stating our assumptions and the basic set up.
First, we assume the existence of a stationary cosmological horizon surrounding the black hole. 
No assumption is made about the asymptotic behavior of the scalar field or the metric.     
We further assume that the spacetime is stationary, axisymmetric, torsion-free and with no naked curvature singularity anywhere in the region of interest, which is the region between the black hole and the cosmological event horizon. The trace of the energy-momentum tensor appearing in the Brans-Dicke field equation of motion may have several components. Firstly, conformally invariant matter fields like  Maxwell, Yang-Mills, conformal scalar or massless fermions, have $T=0$. Each of the remaining components will be assumed to be an ideal fluid with $T_{ab}=\rho u_a u_b +P \left(u_au_b+g_{ab}\right)$, 
where $u_au^a=-1$, $\rho$ and $P=w\rho$ with constant $w$ are, respectively, the energy density and pressure of the fluid.
We also assume that each of the fluid components has positive energy density: $\rho\geq 0$.
For nonrelativistic matter $w\ll1$, while for radiation $w=1/3$. Thus, we shall be interested only in  fluids with $w \leq 1/3$, for which $T=-\rho\left(1-3w\right)\leq 0$ always, including the dark energy-like ones.

As for the geometric set up we are going to use, its details can be found in~\cite{Wald:1984rg}, and in e.g.~\cite{Bhattacharya:2013caa} and references therein. It is motivated by the Kerr-Newman-de Sitter family of spacetimes.

Any stationary and axisymmetric spacetime is endowed with two Killing vector fields $\xi^a$ (timelike) and $\varphi^a$ (spacelike), which generate stationarity and axisymmetry, respectively. We assume that these two Killing vector fields commute,
\begin{eqnarray}
\left[\xi,\varphi\right]^a=\xi^b \partial_b \varphi^a - \varphi^b \partial_b \xi^a = 0.
\label{bdad2}
\end{eqnarray}
This is trivially true when these are coordinate vector fields, i.e. $\xi^a=(\partial_t)^a$ and $\varphi^a=(\partial_{\varphi})^a$. 
We assume that the 2-planes orthogonal to these commuting Killing vector 
fields foliate the space-time~\cite{Wald:1984rg}. Roughly speaking, this means that as we move such a plane along $\xi^a$ and $\varphi^a$ and generate a family of them, those planes do not `twist' or `bend' into the directions of those Killing fields.
The chief technical difference of the stationary axisymmetric spacetime (e.g. Kerr) with that of static (e.g. Schwarzschild)
is the fact that for the former we do not have any timelike Killing vector field, which is orthogonal to a family of spacelike hypersurfaces (e.g. for Kerr, $g_{t\varphi}\neq0$). 

For our convenience, we shall first construct a family of spacelike hypersurfaces, by defining a vector field $\chi_a=\xi_a-(\xi\cdot\varphi  /\varphi\cdot\varphi)\varphi_a$, such that $\chi_a\varphi^a=0$ identically. Then, it turns out that (e.g.~\cite{Bhattacharya:2013caa}, and references therein):
(a) $\chi_a$ is timelike in our region of interest, $\chi_a\chi^a\equiv -\beta^2\leq 0$, and orthogonal to a family of spacelike hypersurfaces,
(b) on any $\beta^2=0$ hypersurface, where $\chi_a$ is null, the function $\xi\cdot\varphi  /\varphi\cdot\varphi$ becomes a constant, and hence $\chi_a$ is a Killing vector field there. 
(c) For our case we have two $\beta^2=0$ hypersurfaces, the smaller one is the black hole and the larger one surrounding it is the cosmological event horizon. 


The form of the metric in this basis is
\begin{eqnarray}
g_{ab}=-\beta^{-2}\chi_a\chi_b+f^{-2}\varphi_a\varphi_b+\gamma_{ab},
\label{bd34}
\end{eqnarray}  
where $f^2$ is the norm of $\varphi^a$ and $\gamma_{ab}$ is the 2-metric on the 2-planes
orthogonal to both $\chi^a$ and $\varphi^a$.
Similarly, the Brans-Dicke scalar $\phi$ is also assumed to be stationary and axisymmetric, i.e. to satisfy 
$\xi^a\nabla_a\phi=0=\varphi^a\nabla_a\phi$,
which implies that $\chi^a\nabla_a\phi=0$ as well.

Now, the first of~(\ref{fielded}) leads to
\begin{eqnarray}
\mathcal{R}=\frac{2\left(4\Lambda -T\right)\omega}{\phi(2\omega+3)}+\frac{\omega}{\phi^2}\left(\nabla_a\phi\right)\left(\nabla^a\phi\right),
\label{R}
\end{eqnarray}
which indicates that if $\phi$ diverges on the horizon, then $\mathcal{R}$ may also be divergent when the derivative of $\phi$
diverges faster than $\phi$, leading to a naked singularity. Also, an infinite $\phi$ will lead to a vanishing effective Newton's `constant' which is unacceptable. Hence, physically interesting solutions should have finite $\phi$ and $(\nabla\phi)^2$.  We shall determine the exact nature of divergence of $\phi$ at the end of this section and will show indeed it gives divergent $\mathcal {R}$. 

{\it We shall show below that solutions with these properties have necessarily constant $\phi$ and exist only for $\omega \to \infty$. In other words, they reduce to solutions of Einstein's general relativity with a positive cosmological constant}. 

Indeed, using $\sqrt{-g}=\beta \sqrt{h}$, where $h$ is the determinant of the 3-metric $h_{ab}$ on the spatial hypersurface orthogonal to $\chi^a$ in~(\ref{bd34}) and $\chi^a\nabla_a\phi=0$, we obtain
\begin{eqnarray}
\nabla_a\nabla^a\phi=\frac{1}{\beta}D_a\left[\beta D^a\phi\right],
\label{bd35}
\end{eqnarray}  
where $D_a$ is the spatial derivative associated with $h_{ab}$.
%
%
%
%
Thus, equation~(\ref{fielded}) for $\phi$ becomes
\begin{eqnarray}
D_a\left(\beta D^a\phi\right)= \frac{\beta\left(T-4\Lambda\right)}{2\omega +3}.
\label{bd38}
\end{eqnarray}  
We multiply this with $e^{\epsilon\phi}$ (with $\epsilon=\pm 1$), and integrate over the spatial hypersurface orthogonal to $\chi^a$ (say, $\Sigma$) between the two horizons to get
%
\begin{eqnarray}
\int_{\partial \Sigma}\!e^{\epsilon\phi}\beta n^a D_a\phi=\!\!\!\int_{\Sigma}\!\beta e^{\epsilon\phi}\left[\epsilon \left(D^a\phi\right)\left(D_a\phi\right)-\frac{4\Lambda-T }{2\omega +3}\right].
\label{bd39}
\end{eqnarray} 
%
The surface integral is evaluated on the two horizons, where $\beta$ vanishes, and $n^a$ is the spacelike unit normal to the horizons. We recall that the scalar field and $\left(D_a\phi\right)\left(D^a\phi\right)$ are bounded on the horizons. Then since $\beta=0$ on $\partial \Sigma$, the left hand side vanishes~\footnote{Clearly, the same conclusion holds in the presence of more than one black-hole horizons.} and we are left with a vanishing integral over $\Sigma$.

For $2\omega+3>0$, we take $\epsilon=-1$ in Eq.~(\ref{bd39}). Given that $D_a\phi$ is spacelike, the inner product $(D_a\phi)(D^a\phi)$ is positive definite, $T\leq 0$ and $\Lambda>0$, the integrand is the sum of two negative-definite terms. Thus, the vanishing of the integral over $\Sigma$ implies $\omega \to \infty$ and $\phi\!=$constant. 
Similarly for $2\omega+3<0$, with a choice of $\epsilon=+1$. {\it In other words, for finite $\omega$ the scalar field should diverge at the horizons}~\cite{Galtsov} Q.E.D. 

If at least one of the horizons is not a true event horizon, e.g. in the case of a normal star or of a time dependent astrophysical black-hole, the left hand side of (\ref{bd39}) will be non-vanishing and the scalar field cannot be argued to be constant. Let us assume that outside such a stellar object, there exists a surface $\cal S$, where the Brans-Dicke field has an extremum. Then by (\ref{bd39}) it becomes obvious that in between $\cal S$ and the cosmological event horizon, $\phi$ is constant and $\omega = \infty$. Now, for $\omega=\infty$ the theory reduces to Einstein gravity, so that $\phi$ is constant inside $\cal S$ as well. Thus, {\it for finite $\omega$ and in the presence of a positive cosmological constant, the field $\phi$ of a stationary non-black hole spacetime must be monotonic. } This monotonicity result will be important in Section IV.


We have concluded that for finite $\omega$, $\phi$ necessarily diverges on the horizons ($\beta^2=0$). {\it We show next that its behavior is $\phi\sim \ln\beta$} there. Let us define $Z_a\equiv \nabla_a\beta^2$. It satisfies $\chi^a Z_a = 0 = \varphi^a Z_a$. Thus, $Z_a$ is one of the  basis vectors spanning $\gamma_{ab}$ in Eq.~(\ref{bd34}). We recall that on any Killing horizon, we have (see e.g.~\cite{Wald:1984rg}, chapter 12, for details)
\begin{eqnarray}
\nabla_a\beta^2=-2\kappa \chi_a,
\label{bd9}
\end{eqnarray}
where the constant $\kappa$, assumed non-zero, is the surface gravity, given by
\begin{eqnarray}
\kappa^2=\frac{\left(\nabla_a\beta^2\right)\left(\nabla^a\beta^2\right)}{4\beta^2}\Bigg\vert_{\beta^2\to 0}.
\label{bd10}
\end{eqnarray}
Let $Z$ be a parameter along the vector field $Z^a$: $Z^a\nabla_a Z=1$. Then using Eq.~(\ref{bd10}), it is clear that infinitesimally close to the surface $\beta^2=0$, we have
\begin{eqnarray}
 \left(\nabla_a\beta^2\right)\left(\nabla^a\beta^2\right)=Z^a\nabla_a\beta^2=\frac{d\beta^2}{dZ}=4\kappa^2\beta^2,
\label{bd11}
\end{eqnarray}
which can be integrated to give
\begin{eqnarray}
Z=\frac{1}{4\kappa^2}\ln \beta^2.
\label{bd12}
\end{eqnarray}
With this, we shall now solve for  the scalar field in an infinitesimal neighborhood of $\beta^2=0$. We note that both $\chi_a$ and $Z_a$ become null here as $\sim\beta^2$, Eq.~(\ref{bd9}). Then we write $(-\det g)^{\frac12}=\beta^2\gamma(x)$, where $\gamma(x)$ is a nonvanishing and well behaved function. Let $\Theta^a$ be a basis orthogonal to $Z^a$. It is easy to see that the part of the Laplacian coming from $\Theta^a$
does not contribute to the near horizon equation of motion. This is essentially the manifestation of the $1+1$-dimensional dynamics close to a Killing horizon. Since we are assuming stationarity, our theory becomes one dimensional. 

Then using Eqs~(\ref{bd11}), (\ref{bd12}) in (\ref{fielded}) it is easy to obtain
%
\begin{eqnarray}
\phi=\!\!\!\int \!\!dZ \frac{4\kappa^2}{\gamma(x)} \left(\int^{x}\!d\beta^2(x') \Gamma(x')\!\right)\! +\!C_1Z\! +\!C_2,
\label{bd13}
\end{eqnarray}
%
with $\Gamma(x)\equiv(T-4\Lambda)\gamma(x')/(2\omega+3)$ and $C_1$, $C_2$ constants.  

As $\beta^2\to0$, $dZ=d\beta^2/4\kappa^2\beta^2$
is unbounded infinitesimally close to the horizon. Since $\Gamma(x)$ is well behaved, the integral in (\ref{bd13}) becomes  
%
\begin{equation}
\int \frac{dZ}{\gamma(x)} \int^{x}\Gamma(x')d\beta^2(x') 
 = \frac{Z}{\gamma(x)}\left(\int^{x}\Gamma(x')d\beta^2(x')\right)_{\beta^2\to 0} \nonumber
\end{equation}
from which one concludes $\phi \sim Z \sim \ln\beta^2$. Substituting this into~(\ref{R}) and using~(\ref{bd11}) shows 
$\mathcal{R}\sim(\beta\ln \beta)^{-2}$ on the Killing horizons.

However, since $\Gamma(x)$ is finite, expanding it in positive powers of $\beta$, it is easy to see that the first term on the right hand side of~(\ref{bd13}) vanishes on the horizon as $\sim \beta^2$. Thus we can make $\phi$ (and hence $\mathcal{R}$) bounded on the horizon by simply setting $C_1=0$. Clearly, even though we do so, the no hair theorem precisely shows that no global smooth field configuration exists which is finite on both the horizons. In Sec. IV, we shall discuss the possibility of having such finite field configuration on one of the horizons, when the other stationary horizon is absent.

\section{Generalization} 
For general $\omega=\omega(\phi)$ the scalar equation (\ref{fielded}) becomes 
\begin{eqnarray}
\left(2\omega(\phi)+3\right)\Box\phi+ (\partial_\phi \omega(\phi))\left(\nabla\phi\right)^2+\left(4\Lambda-T \right)=0
\label{bdn1'}
\end{eqnarray} 
%
It is now convenient to go to the Einstein frame. Following~\cite{review, Sotiriou:2011dz}, we define $\psi$ by $d\psi\equiv\left(2\omega(\phi)+3\right)^{1/2}d\ln\phi$ and ${\cal{G}}_{ab}\equiv\phi g_{ab}$, so that (\ref{bdn1'}) becomes 
\begin{eqnarray}
{\Box}_{\cal G}\psi = U'(\psi)\equiv -\frac{4\Lambda-T}{\phi^2 \sqrt{2\omega(\phi)+3}} +V'(\psi),
\label{bdn2}
\end{eqnarray}
where $\phi=\phi(\psi)$ as a function of $\psi$ and we have added an arbitrary potential $V$. According to e.g. \cite{Bekenstein,Sotiriou:2011dz} if $U''(\psi)\geq 0$ holds, then the no-hair theorem is valid. 
A few comments are in order here: 
(i) For $V=0$, the no-hair theorem is valid for arbitrary $\omega(\phi)$, $\Lambda\geq 0$ and $T\leq 0$. To verify this, we write (\ref{bdn2}) in terms of the spatial derivative $D$, multiply with $e^{-\psi}$ and integrate by parts over the space between the two horizons. Positivity of all terms in the resulting vanishing integral leads to the conclusion that $\phi$ is constant and $\omega(\phi)\to \infty$. (ii) For $\Lambda=0=T$ the no-hair theorem is always valid as long as $V''\geq 0$. However, (iii) once we switch on $\Lambda$ and/or $T$, 
the validity of the theorem requires, as we said, $U''\geq 0$, which restricts $\omega(\phi)$. {\it Thus, unlike the asymptotically flat spacetimes, the presence of $\Lambda>0$ might lead to hairy black holes even for $T=0$ and for some $V''\geq 0$} -- at least we cannot rule out this qualitatively new and interesting possibility. (iv) Replacing $\Lambda$ by generic dark energy with $w<-1/3$, does not modify our conclusions, provided there exists a cosmological horizon in that case as well. However, although $w<-1/3$ implies repulsive effects, which is necessary for the existence of a cosmological horizon~\cite{Gibbons:1977mu}, it is not known to be also sufficient. 

\section{Spherical star solutions}
 Given that for $\omega\to\infty$ the theory (\ref{lg},\ref{fielded}) reduces to general relativity with $\Lambda\neq 0$, it is reasonable for large $\omega$ to search for solutions perturbatively in powers of $1/\omega$ around $\Lambda{\rm CDM}$ in the form
$
\phi=1+\phi^{(1)}/{\omega}+\ldots, \; g_{ab} = g_{ab}^{(0)} + g^{(1)}_{ab}/{\omega}+\ldots, 
$
where the zeroth-order metric $g^{(0)}_{ab}$ is the Schwarzschild-de Sitter (SdS),
%
\begin{eqnarray}
ds^2=- F(r) dt^2+F(r)^{-1} dr^2+ r^2d\Omega^2,
\label{bd1}
\end{eqnarray}
%
with $F(r)\equiv 1-2M/r -\Lambda r^2 /3 $. Eqn. (\ref{bd1}) gives the space-time of a spherical object embedded in the de Sitter universe and admits two Killing horizons at $F(r)\!=\!0$. For $3M\sqrt{\Lambda}\ll 1$, which is the case of interest in physics (even for $M\sim 10^{18} M_{\odot}$, since $\Lambda\sim 10^{-52}{\rm m}^{-2}$), the two horizons are located at
$r_H\simeq 2M, ~r_C\simeq \sqrt{3/\Lambda}$ and are very widely separated. 
The first correction of $\phi$ satisfies~$\Box_0\phi^{(1)}=-2\Lambda$, where $\Box_0$ corresponds to (\ref{bd1}). Using $r_H\ll r_C$, we obtain up to an irrelevant additive constant
\begin{eqnarray}
\phi^{(1)}=(C_1/r_H)\ln(1-r_H/r)+(1-C_1/2 r_C)\ln(1-r/r_C)
+(1+C_1/2r_C)\ln(1+r/r_C),
\label{bd6}
\end{eqnarray}
where $C_1$ is another constant. In accordance with the theorem, 
it exhibits (since $\beta^2=1-2M/r-\Lambda r^2/3$) a $\log-$divergence at both horizons. 
However, since in realistic astrophysical systems either one or both stationary horizons are absent, special cases of (\ref{bd6}) are of physical interest.

(i) Let us assume that the stationary cosmological horizon is absent, in the sense that it is not Killing due to the breaking of de Sitter symmetry e.g. by some time dependent matter field, in which case (\ref{bd1}) and (\ref{bd6}) can be used only up to a certain $r\ll r_C$. To avoid the divergence at $r_H$ we set $C_1=0$. Then
\be
\phi^{(1)} = \ln(1-r^2/r_{ C}^2)\,,
\label{bhsol}
\ee
which is finite, reliable for all $r_H\leq r \ll r_C$ and non-trivial at $r_H$. Thus, black-hole horizon can support $\phi-$hair if the stationary cosmological horizon is absent. {\it This is not possible, neither in $\Lambda=0$ Brans-Dicke theory, nor in $\Lambda\geq 0$ theory with minimally coupled real scalar.} Evidently, this result can further be supported  by the local non-perturbative argument made at the very end of Sec. II. 

(ii) If only the horizon at $r_H$ is absent, e.g. in the interior of a ``star", we choose $C_1=2 r_{\rm C}$ in (\ref{bd6}), so that
%
\begin{eqnarray}
\phi^{(1)}=(2r_{C}/r_{H})\ln(1-r_{H}/r)+2\ln(1+r/r_{C}),
\label{bd7}
\end{eqnarray}
is regular, monotonically increasing, unique and certainly reliable in the regime $r_C/\omega \ll r \leq r_C$. 

The exterior solution (\ref{bd7}), valid near the cosmological horizon, has to be matched to the interior one at the surface of the star. Furthermore, the interior solution is decreasing at the surface of the star. This can be seen by integrating (\ref{bd38}) from $r=0$ to the surface $r=R$. For $T<0$ we obtain $\int_\Sigma D_a(\beta D^a\phi) < 0$, which for spherical configurations regular at $r=0$, implies $d\phi/dr<0$ at $r=R$. The exterior solution, to match smoothly with the interior, must also be decreasing at the star surface. But, any smooth exterior $\phi(r)$ with negative slope at $R$ and positive slope at $r_C$ (Eq.~(\ref{bd7})) has necessarily an extremum in between. This contradicts the non-perturbative monotonicity result for $\phi$, we proved in Section II. Thus, smooth solutions regular both at $r=0$ and at $r=r_C$ can exist only for $\omega=\infty$, they have $\phi\!=$constant throughout, while their exterior is just the Schwarzschild-de Sitter space-time. 

This result requires that perturbation theory is valid only in the vicinity of the cosmological event horizon. Even though this is perfectly reasonable, we may further 
extend this result fully non-perturbatively, also to the stationary axisymmetric case as follows: Let us assume we have a non-perturbative and finite $\phi$, analogue of~(\ref{bd7}). We integrate~(\ref{bd38}) on $\Sigma$, from the star surface up to the cosmological event horizon.
For the exterior solution, we find $\int_{S_0}\beta n^aD_a\phi>0$,
where $S_0$ is the spacelike two surface of the star, this time not a sphere. Likewise, for a regular solution at the centre, we find for the interior, $\int_{S_0}\beta n^aD_a\phi<0$. Due to the axisymmetry, we cannot just pull 
$n^aD_a\phi$ out of the surface integrals. However, since the surface is the same for both the interior and the exterior (guaranteed
from the smoothness of the metric functions), it is clear that respectively for the interior and the exterior, negative and positive values of $n^aD_a\phi$ should dominate. This rules out the possibility of matching everywhere at the star surface. 

Consequently, we conclude that in order to describe, for instance, a star in the context of (\ref{lg}) with finite $\omega$, one has to search for solutions in the absence of a stationary cosmological horizon, in the sense of being valid up to a certain $r\ll r_C$, beyond which it ceases e.g. to be stationary. This will be the implicit assumption we use next. 

(iii) Assume that we restrict our interest to $r_H\ll r \ll r_C$. One possibility~\cite{BDRT} in this case is to set $C_1=-r_H/2$ in (\ref{bd6}), so that $\phi^{(1)}$  coincides for $\Lambda=0$ with the known weak field solution of~\cite{Brans}. The same criterion is used below to fix the integration constants for the metric. Thus,
\begin{eqnarray}
\phi^{(1)}\approx -(1/2)\ln(1-r_H/r)+ \ln(1- r^2/r_C^2).
\label{bd6'}
\end{eqnarray}
This is the natural choice in the region $r_H\ll r\ll r_C$ of its validity, away from both horizons.

Eq.~(\ref{bd6'}) is relevant to study the effect of the Brans-Dicke field on the maximum turn around radius~\cite{Pavlidou:2014aia, Pavlidou:2013zha, Tanoglidis:2014lea} of cosmic structures. 
In its region of validity one may insert (\ref{bd6'}) into Eq.~(2) to obtain to leading order in $1/\omega$ and with $T_{ab}=0$
%
\begin{eqnarray}
G_{ab}+\Lambda (1-2/\omega-\phi^{(1)}/\omega)g_{ab}= 
\frac{1}{\omega}[\nabla_a\phi^{(1)} \nabla_b\phi^{(1)}
-\frac{g_{ab}}{2} (\nabla \phi^{(1)})^2 + \nabla_a\nabla_b \phi^{(1)}]
\label{bdn1}
\end{eqnarray}
%
where metric functions on the right hand side correspond to the background~(\ref{bd1}). 
Considering general stationary, spherical ansatz, $ds^{2}=-f(r)dt^2+h(r)dr^2+r^2d\Omega^2$, we find $f(r)\approx 1-2M(1+1/2\omega)/r-\Lambda(1-1/\omega)r^2/3$,~
$h^{-1}(r)\approx 1-2M(1-1/2\omega)/r -\Lambda(1-2/\omega)r^2/3$. The maximum turn around radius i.e. the root of $f'(r)=0$ is
$R_{\rm ta,max}\approx (3M/\Lambda)^{1/3} (1+1/2\omega)$.
%
According to~\cite{Pavlidou:2013zha} consistency with the astrophysical data for galaxies and galaxy clusters requires that $R_{\rm ta,max}$ is necessarily larger than or about equal to $0.9\, (3M/\Lambda)^{1/3}$. This implies that either $\omega>0$ or $\omega\lesssim -5$.  This conclusion is also supported by a full non-perturbative analysis~\cite{BDRT}. 

\section{Outlook} We have seen that a positive cosmological constant leads to important qualitative changes of the Brans-Dicke phenomenology, for both in the context of black hole physics, as well as in astrophysical applications. For black holes, we not only have $\phi={\rm const.}$ (i.e., the usual statement of no hair theorems~\cite{Hawking:1972qk, Sotiriou:2011dz}), but also $\omega=\infty$, forcing the whole theory to General Relativity. To the best our knowledge, this is the first example where the no hair theorem {\it constrains a theory}. 
Violations of this theorem were also discussed in the context of simple generalizations of the theory, as we have seen, due to  positive $\Lambda$.

Most strikingly, we have seen that there can be no regular star solution in the Brans-Dicke theory in the presence of a stationary cosmological event horizon. Note that such ``star no hair" is not present for asymptotically flat boundary conditions. For in that case the outer boundary integral for Eq.~(\ref{bd38}) at the exterior of a star is located at $r\to \infty$, instead of on a horizon. We can evaluate this integral using the Brans-Dicke weak field solution~\cite{Brans} (i.e. $r\to \infty$ and $\Lambda=0$ in~(\ref{bd6'})). It is negative (as opposed to the case of  the horizon, where it is vanishing) and consequently, there is no ruling out of matching, between the interior and the exterior of the star.

Thus the Brans-Dicke theory with a positive $\Lambda$ can describe physics only when a stationary cosmological event horizon is absent. In other words, for this theory to exist in nature, at very large length scales the spacetime has to deviate from de Sitter, or Einstein's gravity. This is analogous to the so called Vainshtein mechanism~\cite{babichev} for massive gravity theories.  We have reached this conclusion based solely upon some simple geometric arguments. 

With the implicit assumption of the absence of a cosmological event horizon, we further made a phenomenological study based on the observed maximum size vs. mass of the cosmic structures, which led to constraints on the parameter $\omega$. 
Several issues, such as (a) the explicit derivation of the non-perturbative version of the solution~(\ref{bhsol}) carrying $\phi-$hair, (b) the study of specific realizations of the necessary assumption of absence of the stationary cosmological horizon, and (c) the subsequent complete analysis of the predictions of the theory about the maximum size of cosmic structures and their comparison to the data, are worth of further study. 

\textbf{Acknowledgements:}
We acknowledge valuable discussions with C. Charmousis, D. Gal'tsov and C. Skordis. Research implemented under the ``ARISTEIA II" Action of the Operational Program ``Education and Lifelong Learning" and is co-funded by the European Social Fund (ESF) and Greek National Resources. 


\end{document}